# Fragment production in central heavy-ion collisions: reconciling the dominance of dynamics with observed phase transition signals through universal fluctuations[Δ]

J.D.Frankland[1*], A.Chbihi[1], S.Hudan[1], A.Mignon[1], and A.Ono[2]
for the INDRA and ALADIN collaborations:
G.Auger[1], Ch.O.Bacri[3], M.L.Begemann-Blaich[4], N.Bellaize[5], R.Bittinger[4], B.Borderie[3], R.Bougault[5], B.Bouriquet[1], A.Buta[5], J.L.Charvet[6], J.Colin[5], D.Cussol[5], R.Dayras[6], N.De Cesare[7], A.Demeyer[8], D.Doré[6], D.Durand[5], E.Galichet[3,9], E.Gerlic[8], D. Gourio[4], D.Guinet[8], B.Guiot[1], G.Lanzalone[10], Ph.Lautesse[8], F.Lavaud[3], J.L.Laville[1], J.F.Lecolley[5], A.Le Fèvre[4], R.Legrain[6,#], N.Le Neindre[1], O.Lopez[5], U.Lynen[4], W.F.J.Müller[4], L.Nalpas[6], J.Normand[5], H.Orth[4], M.Pârlog[11], P.Pawlowski[3], E.Plagnol[3], M.F.Rivet[3], E.Rosato[7], R.Roy[12], A.Saija[4], C.Sfienti[4], C.Schwarz[4], G.Tabacaru[11], B.Tamain[5], W.Trautmann[4], A.Trczinski[13], K.Turzó[4], E.Vient[5], M.Vigilante[7], C.Volant[6], J.P.Wieleczko[1], B.Zwieglinski[13]

[1]GANIL, CEA et IN2P3-CNRS, B.P. 55027, F-14076 Caen Cedex 05, France, [2]Department of Physics, Tohoku University, Sendai 980-8578 Japan, [3]Institut de Physique Nucléaire, IN2P3-CNRS, F-91406 Orsay Cedex, France, [4]Gesellschaft für Schwerionenforschung mbH, D-64291 Darmstadt, Germany, [5]LPC, IN2P3-CNRS, ISMRA et Université, F-14050 Caen Cedex, France, [6]DAPNIA/SPhN, CEA/Saclay, F-91191 Gif-sur-Yvette Cedex, France, [7]Dip. di Scienze Fisiche e Sez. INFN, Università di Napoli "Federico II", Napoli, Italy, [8]Institut de Physique Nucléaire, IN2P3-CNRS et Université, F-69622 Villeurbanne Cedex, France, [9]Conservatoire Nationale des Arts et Métiers, Paris, France, [10]Laboratorio Nazionale del Sud, Via S. Sofia 44, I-95123 Catania, Italy, [11]Nat. Inst. For Physics and Nuclear Engineering, Bucharest-Magurele, Romania, [12]Laboratoire de Physique Nucléaire, Université Laval, Québec, Canada, [13]Soltan Institute for Nuclear Studies, Pl-00681 Warsaw, Poland
[#]Deceased
(28 February 2002)

**Abstract**
Fragment production in central collisions of Xe+Sn has been systematically studied with the INDRA multidetector from 25 to 150 AMeV. The predominant role of collision dynamics is evidenced in multiple intermediate mass fragment production even at the lowest energies, around the so-called ʽmultifragmentation thresholdʼ For beam energies 50 AMeV and above, a promising agreement with suitably modified Antisymmetrised Molecular Dynamics calculations has been achieved. Intriguingly the same reactions have recently been interpreted as evidence for a liquid-gas phase transition in thermodynamically equilibrated systems. The universal fluctuation theory, thanks to its lack of any equilibrium hypothesis, shows clearly that in all but a tiny minority of carefully-selected central collisions fragment production is incompatible with either critical or phase coexistence behaviour. On the other hand, it does not exclude some similarity with aggregation scenarios such as the lattice-gas or Fisher droplet models.

## 1.INTRODUCTION

The multiple production of intermediate mass fragments, or multifragmentation, observed in heavy ion collisions at bombarding energies above ~20AMeV, has long suggested the tantalising possibility that it may somehow be related to a phase transition of

---

nuclear matter[1]. Mean field calculations which reproduce successfully nuclear ground state properties predict a nuclear matter phase diagram analogous to that of van der Waals fluids[2], with a coexistence region at sub-nuclear densities terminating in a critical point at a temperature of ~18MeV and density ~$\rho_0$/3. One then imagines central heavy ion collisions leading to the formation of initially hot and dense pieces of finite nuclear matter, rapidly expanding and cooling until liquid nuclear droplets condense from the vapour[3,4]. Depending on the initial conditions, this process could be a phase transition of first or second order.

However, it is dangerous to obviate the problem of reaction mechanisms in the study of heavy ion collisions by only considering thermodynamical aspects. Before such studies can be considered as phase transition studies one has to first verify, for example, that the collisions in question do indeed create bulk-excited nuclear matter. This is far from obvious in all situations and remains an open question, with mainly model-dependent answers. Schematically, and at the risk of caricature, transport models derived from semi-classical mean field theory tend to favour the "heated nuclear liquid drop" picture, whereas those based on more-or-less quantal extensions of molecular dynamics generally prefer a rapid, localised formation of fragments during the lifetime of the collision, without any intermediate excited system. The truth probably lies somewhere between the two extremes.

In this contribution we present several recent studies of fragment production in central collisions of Xe+Sn from 25 to 150AMeV bombarding energy, measured with the 4$\pi$ charged particle multidetector INDRA[5]. Taken together, they show that the majority of fragment production is dominated by non-thermal, non-equilibrated processes, and that neither first- nor second-order phase transitions can be invoked for any but a negligible minority of reactions.

## 2. SELECTION OF CENTRAL COLLISIONS

In order to get the least-biased possible information on reaction mechanisms over a wide range of bombarding energies we have used a well-established method for estimating the geometrical impact parameter and therefore the centrality of collisions from some global variable related to the collision violence[6,7]. For this we used the total transverse energy of hydrogen and helium isotopes (LCP or light charged particles), $E_{trans12}$. The experimentally measured distributions of this variable have an almost identical form when rescaled with respect to beam or available energy (see Figure 1). This suggests that $E_{trans12}$ is principally sensitive to the collision geometry.

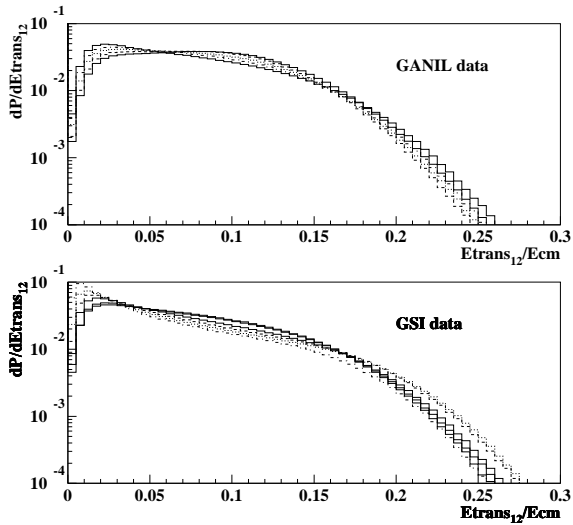

Figure 1: Distributions of total transverse energy of light charged particles measured with INDRA for Xe+Sn collisions from 25-50AMeV (upper panel) and 50-150AMeV (lower panel). The experimental trigger of at least 4 detected charged particles was applied.

In addition, the fact that the INDRA array has an almost 90% efficiency for LCP regardless of centrality makes this variable particularly well-suited to an unbiased event selection. It should be stressed however that this approach is not sufficiently sensitive in order to isolate data samples compatible with the multifragmentation of a well-defined, thermodynamically equilibrated source[8], such as those studied in Refs.[9,10,11]. Our aim here is rather to obtain a *vue d'ensemble* of fragment production in central collisions, not only that which may be attributed to a thermodynamic origin. In the following we therefore studied events belonging to the ~10% (Sections 3-5) or 1% (Section 6) most violent collisions measured by the total transverse energy of light charged particles. These event samples correspond to effective geometrical impact parameters $b<0.3b_{max}$ or $b<0.1b_{max}$, respectively, where $b_{max}$ is the maximum geometrical impact parameter for reactions satisfying the experimental trigger condition of at least 4 charged particles detected.

### 3. OVERVIEW OF FRAGMENT PRODUCTION IN CENTRAL XE+SN COLLISIONS

Let us begin with an overview of fragment production in central collisions from 25 to 150 AMeV. Fragment charge distributions (Figure 2, left) evolve from very broad shapes at 25 AMeV, where residues heavier than projectile or target nuclei are observed, towards an exponential form at the highest energies, indicating total fragmentation of both nuclei. The mean multiplicity of produced fragments (here arbitrarily defined as all products with $Z\geq3$) increases to a maximum of 6 at an incident energy of 65AMeV, and then decreases as the average size of all reaction products becomes smaller with increasing energy of the collisions. Lastly, the profile of fragments' mean centre of mass kinetic energies (Figure 2, right) shows a monotonous increase with their charge suggesting incomplete stopping/transparency or perhaps radial flow, except at 25 AMeV where the heaviest fragments observed have low c.m. energies as they are almost at rest in this frame, as one would expect for residues of fusion-evaporation reactions.

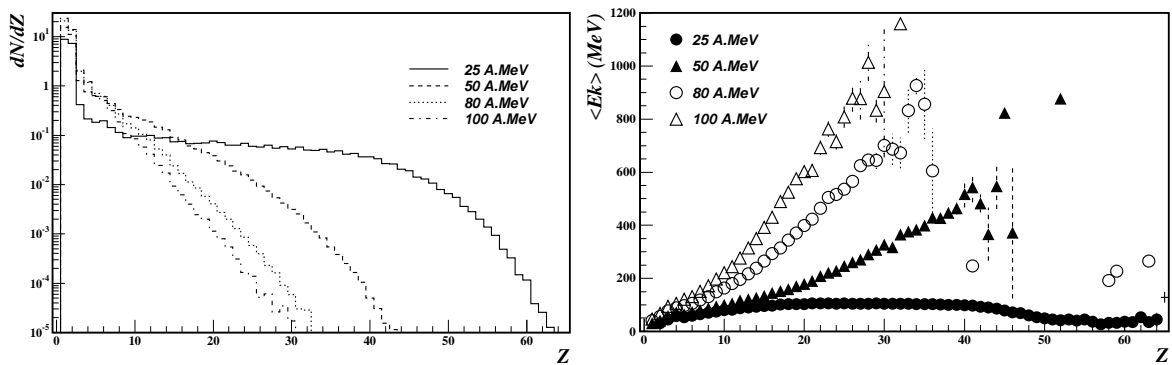

*Figure 2: (left) Charge distributions for central collisions ($b<0.3b_{max}$) of Xe+Sn from 25 to 150AMeV. (right) Mean c.m. kinetic energy of products as a function of detected charge for the same collisions. Vertical bars show estimated statistical errors where larger than the symbols used.*

In the following, we will study the mechanism of fragment production in these collisions in detail, beginning with the lowest energy. Then we will present a global analysis of all the data in order to show what one may discern with certainty (i.e. independently of any hypothesis on reaction mechanism, equilibrium, etc.) about the possible signals of phase transitions in this data.

# 4. HEAVY RESIDUE AND IMF PRODUCTION AROUND THE "MULTIFRAGMENTATION THRESHOLD"

We will begin our detailed study of the reactions with the data for 25 AMeV. This energy is particularly interesting as, for (incomplete) fusion reactions, one would expect a maximum excitation energy of the compound nuclei of 4-5 AMeV, where one expects multifragmentation to become a dominant decay channel of hot nuclei. As shown above, central collisions of Xe+Sn at this energy can produce residues heavier than both projectile and target that recoil at around the centre of mass velocity. To make a detailed study of these reactions, we put an (offline) trigger on the charge and the (laboratory) recoil energy of the heaviest fragment. We observed, for the heaviest residues ($Z_{max} \geq 48$), a clear separation of their measured energy spectra at forward angles $\theta \leq 15°$ into two components, corresponding to high energy projectile-like fragments and low-energy fusion-like residues with small centre of mass velocities (target-like fragments with energies sufficient to pass identification thresholds of ~1 AMeV were only observed at larger polar angles).

Coincident LCP velocity diagrams, shown in Figure 3(a), have well-defined Coulomb circles centred on the c.m. velocity, strongly suggestive of a fusion-evaporation scenario. It is then tempting to associate the coincident IMF yield ($<M_{IMF}> \approx 2$), much higher than can be accounted for by standard statistical decay as modelled in e.g. the GEMINI code[12], with some "new" physics i.e. the onset of multifragmentation due to the increase of the excitation energy of the produced hot nuclei above the threshold for this process.

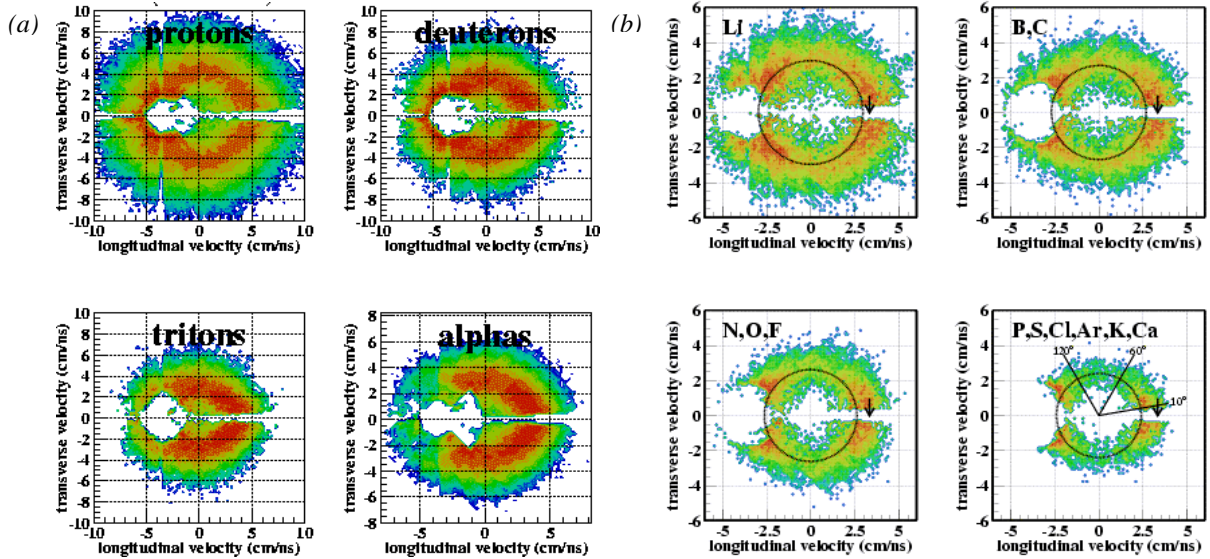

*Figure 3: Invariant velocity diagrams in the c.m. frame for (a) light charged particles and (b) light fragments detected in coincidence with a heavy residue ($Z_{max} \geq 48$) in the 25AMeV reactions.*

However, velocity diagrams for coincident IMF (Figure 3(b)) show a strong anisotropic component focused around the forward and backward beam directions. Thanks to the quasi-complete kinematical reconstruction of these events provided by the INDRA $4\pi$ array, we examined event-by-event the relative angle between IMF velocity vectors with respect to the residue and found that IMF are preferentially emitted in correlated pairs with small relative angles $\approx 30°$ (Figure 4). By performing three-body Coulomb trajectories calculations we were able to test various scenarios for events leading to a heavy residue and two, much smaller, IMF (these account for most of the yield). These calculations clearly exclude the possibility of random isotropic emission of the fragments by the (parent of the) heavy residue. The best agreement with the experimentally observed kinematical correlations is achieved by supposing that the two IMF result from the break-up of some small remnant or spectator of <Z>~15 ≥250 fm/c after the collision (Figure 5).

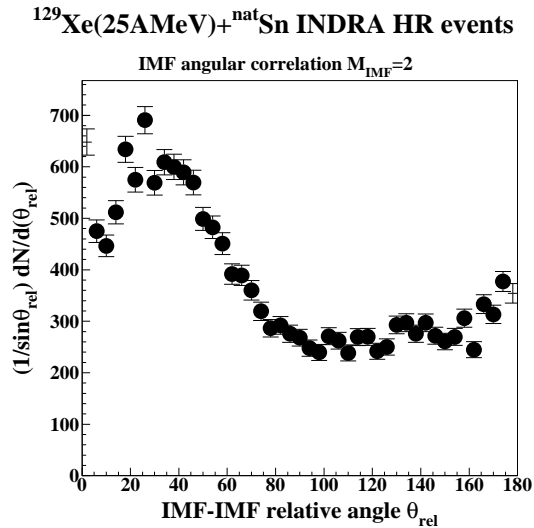

*Figure 4: Distribution of the polar angle between pairs of IMF velocity vectors relative to the heavy residue in central collisions of Xe+Sn at 25AMeV.*

We conclude that the increased yield of IMF in these central collisions is not due to the onset of thermal multifragmentation in a compound nucleus, but rather it is due to reaching the limits of mass and excitation energy that may be reached via a heavy-ion collision. We speculate that the composite system that would have been formed would have been too excited and/or too heavy in order to survive the collision, even after pre-equilibrium emission, and so the reaction proceeds via a two-step binary process: deep inelastic collision with a very large, asymmetric transfer of mass and energy towards one of the two partners, followed by the break-up of the small remainder in the exit channel. As for the other partner, preliminary analysis suggests that a large hot nucleus is formed which decays according to statistical low-energy theory (neutrons, light charged particles, a very small IMF multiplicity and fission, which does not contribute to the observed yield of heavy residues).

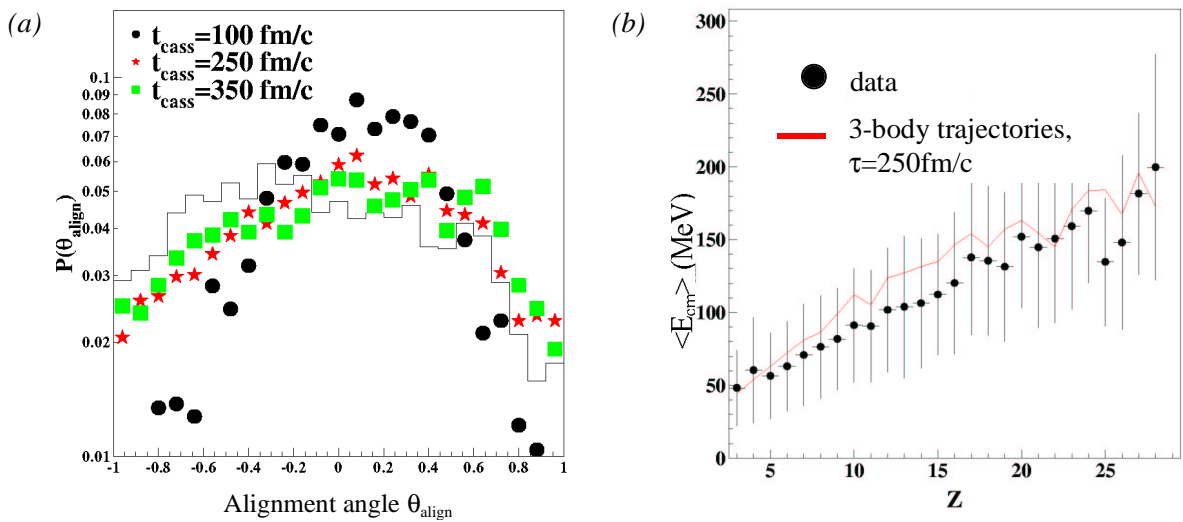

*Figure 5: Comparison of data with 3-body Coulomb trajectories calculations for break-up of projectile/target remnants after formation of a heavy excited nucleus. (a) Determination of break-up time from the alignment of the "break-up axis" and the "recoil axis" (histogram corresponds to data). (b) Mean c.m. kinetic energies of IMF as a function of their charge.*

We would like to insist on the fact that such detailed informations on the reaction mechanisms in these collisions were only obtained thanks to the complete kinematical reconstruction afforded by a 4π detector array. As such apparatus were not available when much of the groundwork was done on "well understood" low energy reactions, we wondered what else might be learnt if such low energy reactions were to be measured with INDRA, apart from establishing at which energy occurs the onset of the fragment production mechanism we observed at 25 AMeV. For this purpose we conducted a new series of experiments at the GANIL facility in Spring 2001 where the Xe+Sn and Xe+Au systems were studied from 8 to 25 AMeV bombarding energy. This data is currently undergoing analysis.

### 5. MULTIFRAGMENTATION AT 50 & 100 AMEV

Let us now examine the evolution of fragment production with increasing bombarding energy. One of the principal characteristics of fragment production observed in central collisions is the anisotropy of the fragment yield[13]. As we showed in the previous section, this anisotropy is present at the onset of multiple fragment production, and it continues to be a feature of reactions with increasing energy (although there may be evidence of increasing isotropy with increasing energy[14]). Such anisotropy suggests a lack of complete stopping even in the most central collisions, as has been observed at higher energy[15]. It is interesting to compare such data with the predictions of dynamical calculations where such effects may be related to nuclear matter properties such as the equation of state, momentum dependent interactions or in-medium cross-sections for nucleon-nucleon scattering.

Among the existing models, the Antisymmetrised Molecular Dynamics [16] features multi-body correlations and stochastic quantum branching processes essential for the description of multifragmentation reactions, in which large fluctuations can exist among the final states of initially similar collisions. In addition, the use of antisymmetrised wave functions ensures the inclusion of the fermionic nature of nuclear dynamics in a consistent manner, whilst nuclear ground states are exceptionally well-described.

We have performed AMD calculations for Xe+Sn collisions at two energies (50 and 100AMeV), in both cases mixing with appropriate weights collisions at all impact parameters up to 4fm. It is only recently that AMD calculations have become feasible for such heavy systems in a reasonable amount of CPU time. Each calculation was carried out up to 300fm/c after the collision, at which time all of the produced fragments were sufficiently separated in the classical (**r**,**p**) phase space in order to be recognised by a minimum spanning tree algorithm. For each nucleus present at this stage of the calculation, its charge, mass, position, momentum, thermal excitation energy and spin were determined. Event by event, these were then used as inputs to a statistical decay code in order to simulate the very much slower decay to the final ground state of the fragments produced, including the Coulomb trajectories of all the nuclei present at each time step up to the detectors. Finally, the asymptotic nuclei were treated with the INDRA software filter in order to take into account effects of the experimental apparatus. More details can be found in[17].

These calculations account quite well for the observed fragment production in central collisions at both energies, including fragment multiplicities and charge distributions (within the limits imposed by the rather low calculated statistics), angular distributions and kinetic energies (Figure 6). At 50AMeV the good agreement with the yields of the heaviest fragments was achieved by inclusion of the wave packet shrinking effect, as well as wave packet diffusion, in the quantum branching process so that the single particle dynamics reproduce the prediction by mean field models more precisely than before[18].

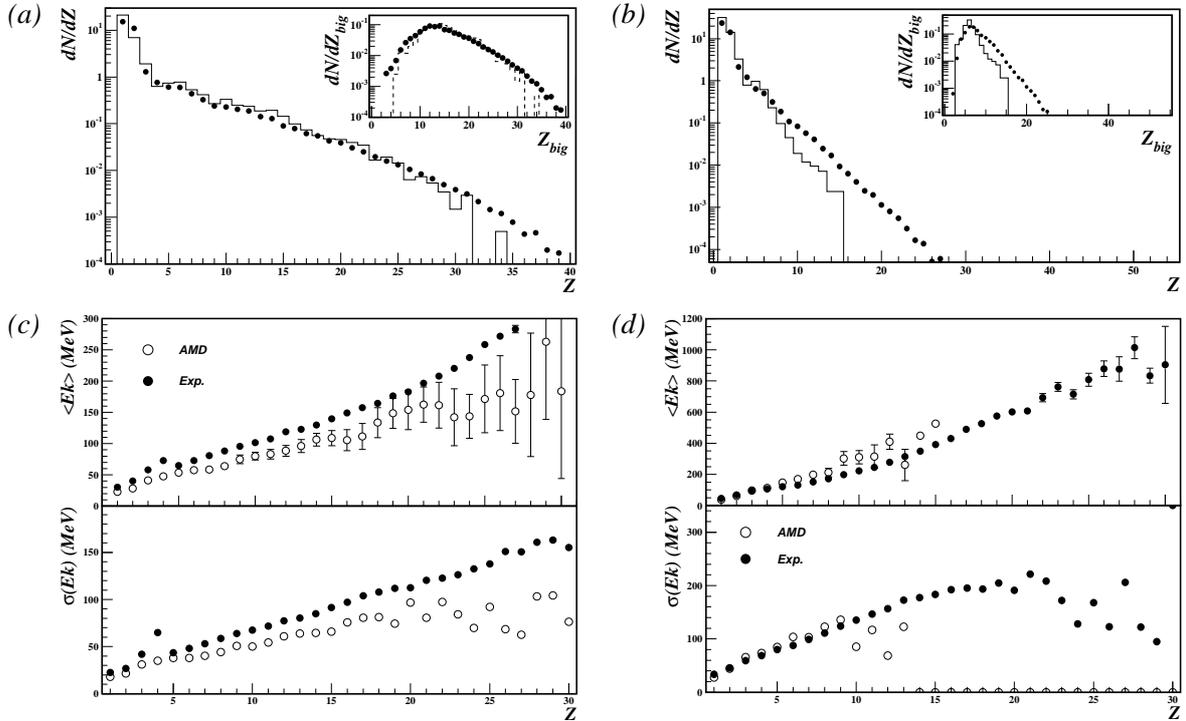

*Figure 6: Comparison of data with AMD calculations for central collisions of Xe+Sn at 50 and 100AMeV. See text for details. (a) Charge distributions at 50AMeV. (b) Charge distributions at 100AMeV. (c) Mean and standard deviation of c.m. kinetic energy distributions of fragments as a function of charge Z at 50AMeV. (d) Mean and standard deviation of c.m. kinetic energy distributions of fragments as a function of charge Z at 100AMeV.*

Based on the success of the model in reproducing data, one may try to infer from it some experimentally inaccessible information about the mechanism responsible for fragment production in these collisions. Although fragments are completely separated in phase space only after 300fm/c, in a typical central collision both projectile and target have completely fragmented in the time between contact and the moment at which they would have reseparated. This can be seen by visualising the positions of the nucleon wave packets in the calculation at each time step. The time scale for fragment production in this case is ~100fm/c at 50AMeV and less at 100AMeV. This image is difficult to reconcile with the formation of an intermediate excited source decaying in a subsequent independent step by multifragmentation: if such a system were to exist, then the timescales for its formation and de-excitation would be the same.

This observation is confirmed in a quantitative manner by the degree of mixing between projectile and target nucleons in each fragment. Fragments produced by a completely equilibrated fused system would be composed of a 50-50 mix (on average) of projectile and target nucleons, independent of velocity. On a scatter plot of mixing versus fragment velocity one would expect the data to lie in a circular region centred on c.m. velocity and 50% mixing. The calculated yield for fragments with different proportions of projectile and target nucleons, on the other hand, shows a strong dependence on the velocity of the fragment (Figure 7). This correlation is the same for all fragment masses and indeed for all impact parameters[17]. Near to the projectile (target) velocity, fragments are almost 100% (0%) composed of projectile nucleons, while fragments close to the centre of mass have the greatest degree of mixing (50%). Similar results were found in QMD calculations for the same system (Xe+Sn at 50AMeV)[19].

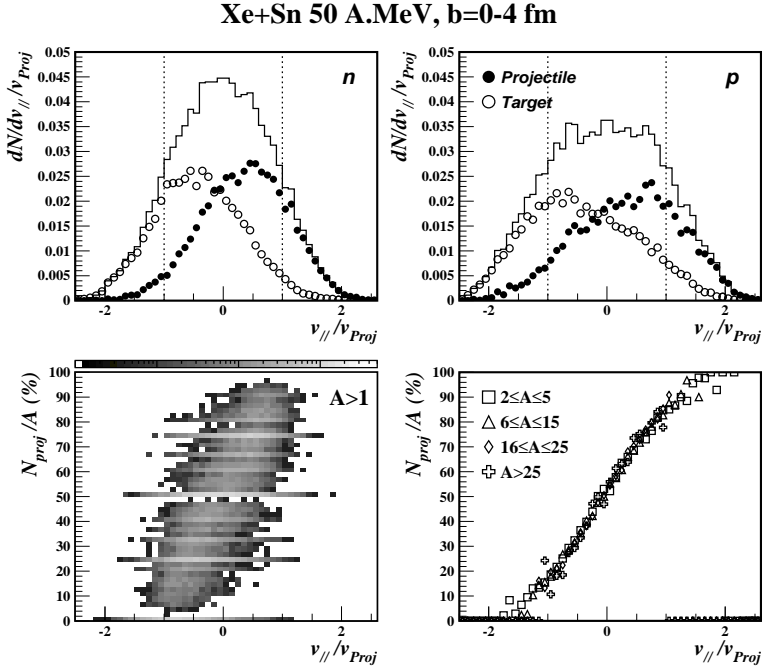

*Figure 7: Results of AMD calculations for central Xe+Sn collisions at 50AMeV after 300fm/c. (top row) velocity distributions for free nucleons showing the decomposition into projectile and target contributions. (bottom left) scatter plot showing the correlation between cluster composition (percentage of nucleons originating in the projectile) and velocity. (bottom right) the mean composition of each cluster as a function of its velocity and mass.*

This is exactly what one expects from linear momentum conservation if fragments rapidly form by aggregation of nucleons from projectile and target as they pass each other on straight-line trajectories. It is perhaps surprising that a model that treats nucleons as quantal objects should arrive at this result: however the substantial width of the observed correlation about this mean behaviour probably reflects the non-classical nature of the trajectories and the Fermi motion of the nucleons in the two colliding nuclei. This correlation is therefore a quantitative measure of the non-equilibration of fragment production in the AMD calculations.

Of course, such information is of little interest if it cannot be subjected to experimental verification. The FOPI collaboration has recently shown that it is possible to 'tag' projectile and target nucleons by studying collisions of pairs of nuclei with different N/Z ratios[15]. Then one may estimate the degree of mixing between projectile and target nucleons in each fragment from its measured N/Z ratio. At GANIL last year, an exploratory experiment was performed using INDRA to study $^{124,136}$Xe+$^{112,124}$Sn reactions at 32 and 45AMeV. The addition of a few high-resolution thin silicon detectors to the INDRA array for this experiment should allow good isotopic identification up to Z=8-9 in the forward rings. Then it may be possible to obtain some quantitative information about mixing in central collisions at these energies. However one has to be careful in such studies to avoid spurious effects on isotope yields such as e.g. binding energy differences between isotopes with different neutron-to-proton ratios, and side-feeding corrections due to secondary decay of hot fragments[20]. In order to minimise such difficulties a projectile and target combination with the largest possible N/Z difference is probably best. Therefore the definitive experimental test may have to wait for 2$^{nd}$ generation radioactive beam facilities capable of delivering beams of near-dripline heavy ions of mass ~100-150 with intensities of $10^4$-$10^6$pps., and 3$^{rd}$ generation 4$\pi$ arrays with charge and mass identification over a wide range of isotopes, up to Z~40.

## 6. UNIVERSAL FLUCTUATIONS AND THE QUESTION OF CRITICALITY

In the data we have presented above there are clear signals that most fragment production in central collisions cannot be reduced to the statistical decay of thermally equilibrated systems. This is not sufficient, however, to exclude the possibility that fragment production may be related to a critical process such as a phase transition, which may occur just as well in off-equilibrium systems as in equilibrated ones. The importance of our experimental observations for the question of the phase transition is that almost all the proposed theoretical signals are based on the same fundamental hypothesis of a statistically equilibrated system. When data are not compatible with such a hypothesis the meaning of these signals is not clear.

In a first approach to the question of the relation between multifragmentation and criticality, we have tried to firmly establish the most that can be said about a possible phase transition in intermediate energy heavy-ion collisions in the least hypothesis-dependent way possible. To do this the theory of universal order parameter fluctuations in finite systems has been applied to a large set of data obtained with the INDRA multidetector[21],[22].

Universal scaling laws of fluctuations (the $\Delta$-scaling laws) can be derived for equilibrium and off-equilibrium systems when combined with the finite-size scaling analysis[23]. In any system in which the second-order critical behaviour can be identified, the relation between order parameter, criticality and scaling law of fluctuations has been established and the relation between the scaling function and the critical exponents has been found. In layman's terms the way in which the fluctuations of an extensive observable scale with system size can be expressed as $\sigma \sim <m>^\Delta$ with $0.5 \leq \Delta \leq 1$. If the probability distributions of the observable in question for different system sizes $N$ are re-expressed as

$$\langle m \rangle^\Delta P_N[m] \equiv \Phi(z_{(\Delta)}) \equiv \Phi\left(\frac{m-m^*}{\langle m \rangle^\Delta}\right) \tag{1}$$

where $m^*$ is the most probable value of $m$, then all probability distributions collapse to a single universal scaling function $\Phi$ for a given value of the scaling exponent $\Delta$.

The results of applying this procedure to data for central collisions of Xe+Sn are shown in Figure 8. The distributions of the size of the largest fragment detected in each event[*] collapse to a single curve over two distinct bombarding energy ranges, with two different values of $\Delta$. At low energies (E<45AMeV) the scaling with $\Delta$=1/2 is observed, meaning that fluctuations are Poissonian, as one would expect for independent stochastic processes. At higher energies (E>45AMeV) a $\Delta$=1 scaling is observed indicating the onset of large fluctuations. It should be noted that the bombarding energy at which the transition occurs depends on the selection of events studied: it was shown in Ref. 21 that it increases with decreasing centrality. For the same data, fragment multiplicities obey a single $\Delta$=1/2 scaling law with a gaussian scaling function from 25 to 150AMeV.

---

[*] In order to minimise the possibility that in any event a larger fragment may have gone undetected, we required that at least 80% of the total charge of the colliding nuclei was measured in the events included in this analysis.

Similar scaling behaviour is also observed for many other systems with widely different masses (Figure 9). For the heaviest system (Au+Au, 394 nucleons) a Δ=1 scaling law holds for all studied bombarding energies (40-100AMeV). The two lightest systems (Ar+KCl, 73 nucleons and Ar+Ni, 94 nucleons) present a Δ=1/2 scaling up to the highest measured bombarding energies (with perhaps the beginnings of a transition towards the Δ=1 regime). An intermediate system, lighter than Xe+Sn (Ni+Ni, 116 nucleons), shows behaviour intermediate between the two extremes, with both scaling regimes being present and the changeover occurring at a beam energy slightly higher than for Xe+Sn.

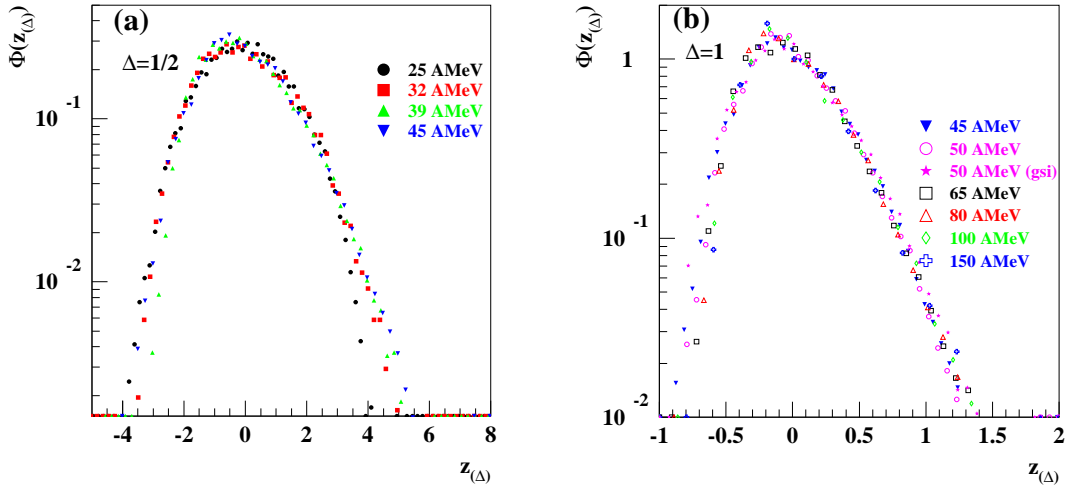

*Figure 8: Probability distributions of the charge of the largest fragment, $Z_{max}$, for central collisions ($b_{red}<0.1$) of Xe+Sn from 25 to 150 AMeV, plotted in the variables of the (a) Δ=1/2 and (b) Δ=1 scaling.*

To summarise the Δ-scaling behaviour we have observed in data for central collisions studied so far:
1. Fragment multiplicities have gaussian scaling functions and small fluctuations ($\sigma \sim \langle m \rangle^{1/2}$) for all energies and all systems.
2. The largest fragment of each event exhibits different scaling behaviour depending on the entrance channel and the centrality of the reaction:
    a. At the lowest energies, in the lightest systems and/or in the least central collisions, the largest fragment obeys a Δ=1/2 scaling law.
    b. At the highest energies, in the heaviest systems and/or in the most central collisions, the largest fragment obeys a Δ=1 scaling law.
    c. The incident energy at which the change from one scaling regime to the other takes place decreases for more central collisions and for heavier systems. It also exhibits some dependence on entrance channel asymmetry (the transition energy for Ar+Ni does not follow the systematics for the symmetric systems).

A possible interpretation of such behaviour is the following. There are two generic families of fragment production scenarios for which the second-order phase transition has been identified, with two different order parameters. These are the fragment multiplicity (*fragmentation scenarios*, e.g. fragmentation-inactivation-binary model[24]) and the size of the largest fragment (*aggregation scenarios*, e.g. percolation, lattice-gas or Fisher drop model[25]). In each case, the order parameter exhibits a change of Δ-scaling regime when a suitable control parameter (e.g. available energy, temperature) is varied, taking the system from an ordered (or subcritical) to a disordered (supercritical) phase. The Δ-scaling analysis therefore allows to exclude the possibility of data being compatible with a fragmentation scenario, while suggesting a certain similarity with models of the aggregation type. The observed behaviour is however not sufficient to claim that any kind of phase transition has actually been observed and is only a necessary condition.

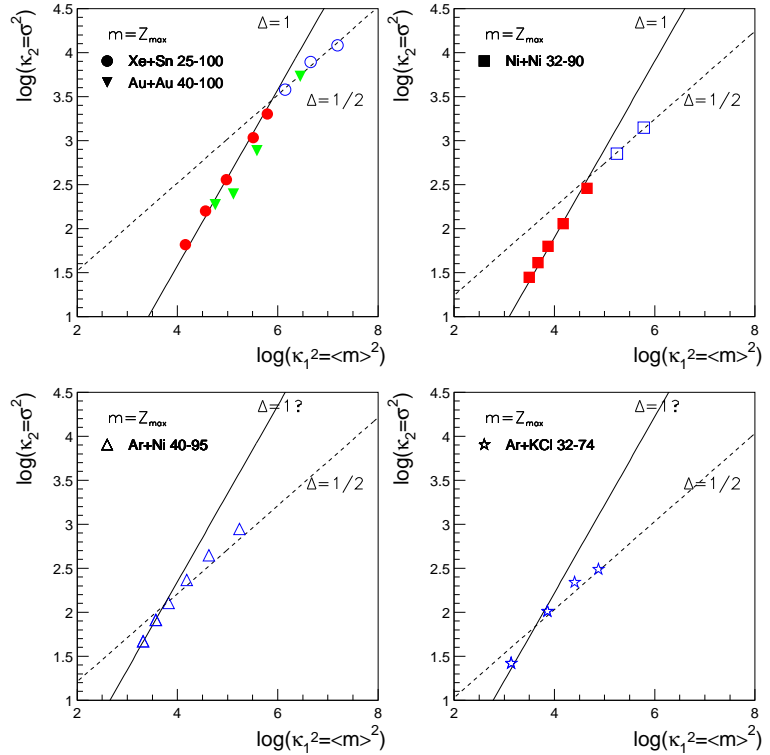

*Figure 9: Compilation of Δ-scaling results concerning the largest fragments in central collisions ($b_{red}<0.1$) of Au+Au, Xe+Sn, Ni+Ni, Ar+Ni and Ar+KCl from 25 to 150AMeV. For each system the lowest bombarding energy (largest mean $Z_{max}$) corresponds to the rightmost point on the plot, the incident energy increases going towards the left. The lines marked Δ=1/2 and Δ=1 are to guide the eye. Open (closed) symbols represent systems obeying a Δ=1/2 (Δ=1) scaling law.*

Further information can be obtained by studying in detail the form of the scaling function, Φ:
(i) In critical systems the order parameter obeys the Δ=1 scaling law with the large $z_{(\Delta)}$-tail of the scaling function falling off like $\Phi \sim \exp(-z_{(\Delta)}^{\nu})$ with ν>2; near the critical point, ν=2 and 1/2≤Δ≤1. In all data studied so far, neither the large exponent tail nor the "cross-over" scaling has been observed. We therefore conclude that none of the studied reactions involves a passage through or near to a critical point.
(ii) At coexistence (i.e. for a 1$^{st}$ order liquid-gas phase transition) one expects the "double-humped" or bimodal form of the scaling function for the order parameter. This behaviour has not been observed in this study. We therefore exclude the possibility that the fragment production studied here is due to liquid-gas coexistence.

Let us recall that for central Xe+Sn collisions between 32 and 50AMeV, the negative heat capacity[10] and spinodal decomposition[11] signals, both associated with exploration of the coexistence region, have been observed by the INDRA collaboration. However, as we remarked earlier, those signals are seen for a far more restrictive sample of central collisions than those studied here, namely those reactions that are compatible with the multifragmentation of a well-defined, thermodynamically equilibrated source. In the present, more general case of "central collisions" defined quite arbitrarily in terms of collision violence, it is clear that such an hypothesis is not applicable and equilibrated thermal fragment production is not the dominant process.

| System | Total mass | Bombarding energies studied (AMeV) | $\Delta=1/2 \tilde{O} \Delta=1$ energy (AMeV) |
|---|---|---|---|
| Ar+KCl | 73 | 32, 40, 52, 74 | >74 ? |
| Ar+Ni | 94 | 40, 52, 63, 74, 84, 95 | >95 ? |
| Ni+Ni | 116 | 32, 40, 52, 64, 74, 82, 90 | 52 |
| Xe+Sn | 248 | 25, 32, 39, 45, 50, 65, 80, 100, 150 | 45 |
| Au+Au | 394 | 40, 60, 80, 100 | <40 ? |

*Table 1: Summary of systems studied in Figure 9.*

## 7. CONCLUSIONS

We studied central collisions of Xe+Sn with the INDRA charged particle multidetector over a wide range of bombarding energies from 25 to 150AMeV. The bombarding energy-independence of the distribution of the total transverse energy of light charged particles was used to define in an equivalent manner for all energies a selection of events corresponding to the most violent (most central) collisions. For these samples of events there is strong evidence that the majority of the observed intermediate mass fragment production is a direct consequence of the collision dynamics and not compatible with the thermal decay of equilibrated excited nuclear matter. In this case, the application of thermodynamics to multifragmentation data is not a valid method of obtaining information on e.g. the nuclear equation of state. Such information can only be inferred from comparisons with dynamical calculations using some parameterisation of the nuclear interaction. Comparisons with Antisymmetrised Molecular Dynamics suggest that this model can give a unified coherent description of fragment production, at least for energies 50AMeV and above.

We showed that pertinent information can be obtained on possible criticality in multifragmentation data without any thermodynamic hypothesis using universal fluctuation theory. The application not only to the Xe+Sn system but to a wide range of systems measured with INDRA showed that while fragment production in the majority of central collisions may be compatible with aggregation scenarios such as the percolation or Fisher droplet models etc., it shows no sign of a passage either in the vicinity of a critical point or through the coexistence region. Such phenomena, if they exist, must be limited to a tiny minority of events.

*Acknowledgements*

Two of us (A.C. and J.D.F.) would like to thank J.B.Natowitz, R.T.de Souza, and M.Ploszajczak for fruitful and stimulating discussions. One of us (J.D.F.) would like to thank I.Iori and A.Moroni for their hospitality, and W.G.Lynch, Z.Majka, and J.Baudot for invaluable technical assistance that allowed this talk to be given in the best possible conditions.

# REFERENCES

[1] For a recent and thorough review, see for example S.Das Gupta, A.Z.Mekjian and M.B.Tsang, nucl-th/0009033 (submitted to Adv. Nucl. Phys.)

[2] G.Sauer et al., Nucl. Phys. A264(1976)221

[3] G.F.Bertsch and P.J.Siemens, Phys. Lett. 126B(1983)9

[4] J.D.Frankland et al. (INDRA collaboration), Nucl. Phys. A689(2001)940

[5] J.Pouthas et al., Nucl. Inst. And Meth. A357(1995)418 and A593(1996)222

[6] C.Cavata et al., Phys. Rev. C42(1990)1760

[7] J.Lukasik et al. (INDRA collaboration), Phys. Rev. C55(1997)1906

[8] J.D.Frankland et al. (INDRA collaboration), Nucl. Phys. A689(2001)905

[9] R.Bougault et al. (INDRA collaboration), in *XL Int. Wint. Meet. On Nuclear Physics*, Bormio (Italy), edited by I.Iori, Ricerca scientifica ed educazione permanente, 2002

[10] M.D'Agostino et al., Nucl. Phys. A699(2002)795-818

[11] B.Borderie et al. (INDRA collaboration), Phys. Rev. Lett. 86(2001)3252 and in *Proc. XXXIX Int. Wint. Meet. On Nuclear Physics,* Bormio (Italy), edited by I.Iori, Ricerca scientifica ed educazione permanente (nucl-ex/0106007)

[12] R.J.Charity et al., Nucl. Phys. A483(1988)371

[13] A.Le Fèvre, M.Ploszajczak and V.D.Toneev, Phys. Rev. C60(1999)051602; B.Bouriquet et al. (INDRA collaboration), in *Proc. XXXIX Int. Wint. Meet. On Nuclear Physics,* Bormio (Italy), edited by I.Iori, Ricerca scientifica ed educazione permanente, 2001

[14] F.Lavaud, Thèse de doctorat de l'Université Louis Pasteur Strasbourg I, IPNO-T01-06 (2001)

[15] F.Rami et al. (FOPI collaboration), Phys. Rev. Lett. 84(2000)1120

[16] A.Ono, H.Horiuchi, Toshiki Maruyama and A.Ohnishi, Prog. Theor. Phys. 87(1992)1185; A.Ono and H.Horiuchi, Phys. Rev. C53(1996)2958; A.Ono, Phys. Rev. C59(1999)853

[17] S.Hudan, Thèse de doctorat de l'Université de Caen, GANIL T 01 07 (2001) and in preparation

[18] A.Ono, S.Hudan, A.Chbihi and J.D.Frankland, submitted to Phys. Rev. C

[19] O.Tirel, Thèse de doctorat de l'Université de Caen, GANIL T 98 02 (1998)

[20] H.Xi et al., Phys. Rev. C59(1999)1567

[21] R.Botet et al., Phys. Rev. Lett. 86(2001)3514

[22] J.D.Frankland et al. (INDRA collaboration), in *Proc. Int. Work. On Multifragmentation and related topics*, Nov. 2001, Catania (Italy) (nucl-ex/0201020)

[23] R.Botet and M.Ploszajczak, Nucl. Phys. B (Proc. Suppl.) 92(2001)101-113 and references therein

[24] R.M.Ziff and E.D.McGrady, J.Phys. A18, 3027 (1985); E.D.McGrady and R.M.Ziff, Phys. Rev. Lett. 58, 892 (1987)

[25] M.E.Fisher, Physics 3(1967)255